
\magnification=1200
\output={\plainoutput}

\newcount\pagenumber
\newcount\questionnumber
\newcount\sectionnumber
\newcount\equationnumber
\newcount\referencenumber

\def\ifundefined#1{\expandafter\ifx\csname#1\endcsname\relax}
\def\docref#1{\ifundefined{#1} {\bf ?.?}\message{#1 not yet defined,}
\else \csname#1\endcsname \fi}

\newread\bib
\newcount\linecount
\newcount\citecount
\newcount\localauthorcount
\def\article{\def\eqlabel##1{\edef##1{\the\sectionnumber.\the\equationnumber}}
\def\seclabel##1{\edef##1{\the\sectionnumber}}
\def\feqlabel##1{\ifnum\passcount=1
\immediate\write\crossrefsout{\relax}  
\immediate\write\crossrefsout{\def\string##1{\the\sectionnumber.
\the\equationnumber}}\else \fi }
\def\fseclabel##1{\ifnum\passcount=1
\immediate\write\crossrefsout{\relax}   
\immediate\write\crossrefsout{\def\string##1{\the\sectionnumber}}\else\fi}
\def\cite##1{\immediate\openin\bib=bib.tex\global\citecount=##1
\global\linecount=0{\loop\ifnum\linecount<\citecount \read\bib
to\temp \global\advance\linecount by 1\repeat\temp}\immediate\closein\bib}
\def\docite##1 auth ##2 title ##3 jour ##4 vol ##5 pages ##6 year ##7{
\par\noindent\item{\bf\the\referencenumber .}
 ##2, ##3, ##4, {\bf ##5}, ##6,
(##7).\par\vskip-0.8\baselineskip\noindent{
\global\advance\referencenumber by1}}
\def\dobkcite##1 auth ##2 title ##3 publisher ##4 year ##5{
\par\noindent\item{\bf\the\referencenumber .}
 ##2, {\it ##3}, ##4, (##5).
\par\vskip-0.8\baselineskip\noindent{\global\advance\referencenumber by1}}
\def\doconfcite##1 auth ##2 title ##3 conftitle ##4 editor ##5 publisher ##6
year ##7{
\par\noindent\item{\bf\the\referencenumber .}
##2, {\it ##3}, ##4,  {edited by: ##5}, ##6, (##7).
\par\vskip-0.8\baselineskip\noindent{\global\advance\referencenumber by1}}}

\def\beginsection #1
 {{\global\advance\sectionnumber by1}\equationnumber=1
\par\vskip 0.8\baselineskip plus 0.8\baselineskip
 minus 0.8\baselineskip
\noindent$\S$ {\bf \the\sectionnumber . #1}
\par\penalty 10000\vskip 0.6\baselineskip plus 0.8\baselineskip
minus 0.6\baselineskip \noindent}

\def\no{\eqno(\the\sectionnumber
.\the\equationnumber){\global\advance\equationnumber by1}}

\def\beginref #1 {\par\vskip 2.4 pt\noindent\item{\bf\the\referencenumber .}
\noindent #1\par\vskip 2.4 pt\noindent{\global\advance\referencenumber by1}}

\def\ref #1{{\bf [#1]}}

\def\specialbar#1{\setbox1=\hbox{$\scriptstyle #1$}
\setbox2=\vbox{\hrule width 0.8\wd1}
\raise0.5\ht1\hbox{$\scriptstyle{\lower\dp1\box2}\atop\box1$}}
\def\specialbar#1{\setbox1=\hbox{$\scriptstyle #1$}
\setbox2=\vbox{\hrule width 0.8\wd1}
\vbox{\hbox{$\,$\lower\ht1\box2}\hbox{$\,$\raise\ht1\box1}}}
\def\dstar{d^*}
\def\tr{\hbox{tr}\,}
\def\charav#1{\left\langle\chi^{#1}\right\rangle_p}
\article
\centerline{\bf Ray-Singer Torsion, Topological field theories and the
Riemann zeta
function at $s=3$.}
\vskip2\baselineskip
\centerline{by}
\vskip2\baselineskip
\centerline{Charles Nash and D. J. O' Connor}
\par\vskip\baselineskip
\noindent
Department of Mathematical Physics\hfill School of Theoretical Physics
\par\noindent
St. Patrick's College\hfill              Dublin Institute for Advanced
Studies
\par\noindent
Maynooth\hfill                {\it and}\hfill  10 Burlington Road
\par\noindent
Ireland\hfill                            Dublin 4
\par\noindent
\null\hfill                              Ireland
\par\vskip3\baselineskip
\noindent
{{\bf Abstract}: Starting with topological field theories we investigate the
Ray-Singer
analytic torsion in three dimensions. For the lens Spaces $L(p;q)$ an explicit
analytic continuation of the appropriate zeta functions is contructed and
implemented. Among the results obtained are closed formulae for the
individual determinants involved, the large $p$ behaviour of the
determinants and the torsion, as well as an infinite set of distinct
formulae for $\zeta(3)$: the ordinary Riemann zeta function evaluated at $s=3$.
 The torsion turns out to be trivial for the cases $L(6,1)$, $L((10,3)$
and $L(12,5)$ and is, in general, greater than unity for large $p$ and
less than unity for a finite number of $p$ and $q$.}
\beginsection{Introduction}
The torsion studied in this paper has its origins in the 1930's, cf.
Franz \ref{1}, where it was combinatorially defined and used to
distinguish various lens spaces from one another. Given a manifold $M$  and a
representation of
its fundamental group $\pi_1(M)$ in a flat bundle $E$, this Reidemeister-Franz
torsion is a real number which is defined as a particular product of ratio's
of volume elements $V^i$ constructed from  the cohomology groups $H^i(M;E)$.
\par
Since volume elements are essentially determinants then, for any alternative
definition of a determinant, an alternative definition of the torsion can
be given. Now if one uses de Rham cohomology to compute $H^i(M;E)$ then
these determinants become determinants of Laplacians $\Delta^E_p$ on $p$-forms
with coefficients in $E$. But zeta functions for elliptic operators can
be used to give finite values to such infinite dimensional determinants and
so an analytic definition of the torsion results and this is the analytic
torsion of Ray and Singer \ref{2,3,4} given in the 1970's;
furthermore this torsion was proved by them to be independent of the
Riemannian metric used to define the Laplacian's  $\Delta^E_p$.
\par
This analytic torsion coincided, for the case of lens spaces, with the
combinatorially defined Reidemeister-Franz torsion. Finally Cheeger and
M\"uller \ref{5,6} independently proved that the analytic Ray-Singer
torsion coincides with the combinatorial Reidemeister-Franz torsion
in all cases.
\par
Infinite dimensional determinants also occur naturally in quantum field
theories when computing correlation functions and partition functions.
In 1978 Schwarz \ref{7} showed how to construct a quantum field theory
on a manfold $M$ whose partition function is a power of the Ray-Singer
torsion on $M$.
\par
Schwarz's construction uses an Abelian gauge theory but in three
dimensions a non-Abelian gauge theory---the $SU(2)$ Chern-Simons
theory---can be constructed and has deep and important properties established
by
Witten in 1988: Its partition
function is the Witten invariant for the three manifold $M$ and the
correlation functions of Wilson loops give the Jones polynomial invariant for
the link determined by the Wilson loops---cf. \ref{8,9}. Finally the weak
coupling limit of the partition function is a  power of the
Ray-Singer torsion.
\par
We shall be concerned here with the special situation of three dimensions and
with the case where the three manifold $M$ is a lens space.
In the next two sections we describe the precise setting and the
analytic continuation while the final section contains our concluding remarks.
\beginsection{Topological field theories, analytic torsion and lens spaces}
Quantum field theories of the type alluded to in the previous section
are usually referred to as topological quantum field theories or simply
topological field theories.
\par
It turns out that more than one  topological field theory can be used to
give the torsion, for an excellent review of this question cf.
Birmingham et al. \ref{10}. For example one can take the action
$$S[\omega]=\int_{M}\omega_{n}d\omega_{n},\qquad \dim M=2n+1\no$$
where $\omega_n$ is an $n$-form. The partition function is then
$$Z[M]=\int {\cal D}\omega\mu[\omega]\exp[-S[\omega]] \no$$
$S[\omega] $ has a gauge invariance whereby
$S[\omega]=S[\omega+d\lambda]$ and therefore to define the partition
function it is necessary to integrate over only inequivalent field
configurations. The measure ${\cal D}\omega\mu[\omega]$ thus contains
functional delta functions which constrain the integration and play
the role of gauge fixing, together with their associated
determinants. This measure can be constructed using, for example,
 the Batalin Vilkovisky BRST construction \ref{11,12}. We wish to devote
more space here to lens spaces and the computation of their torsion  and so
we turn to that now.
\par
To define the Ray-Singer torsion, or simply torsion, we take a closed compact
Riemannian
manifold $M$ over which we have  a flat bundle $E$. Let $M$ have a
non-trivial fundamental group  $\pi_1(M)$
which is represented on $E$---this latter property arises very naturally
in the physical gauge theory context where it corresponds simply to the
space of flat connections all of whose content resides in their
holonomy---In any case the torsion is then the real number $T(M,E)$ where
$$\ln T(M,E)=\sum_0^n(-1)^qq\ln \det\Delta^E_q,\quad n=\dim M\no$$
The metric independence of the torsion requires that we  assume, in the above
definition, that the cohomology ring
$H^*(M;E)$ is trivial; this means that the Laplacians $\Delta^E_q$ have
empty kernels and so are strictly positive definite. Given this fact one
may use zeta functions to define  $ \det\Delta^E_q$ in the standard way.
Recall that if $P$ is a positive  elliptic differential or
pseudo-differential operator with spectrum
$\{\mu_n\}$ and degeneracies $\Gamma_n$ then its associated zeta
function $\zeta_P(s)$ is a meromorphic function of $s$, regular at
$s=0$, which is given by
$$\zeta_P(s)=\sum_{\mu_n}{\Gamma_n\over\mu_n^s}\no$$
and its determinant $\det P$ is defined by
$$\ln\det P=-\left.{d\zeta_P(s)\over ds}\right\vert_{s=0}\no$$
Using this we have
$$\ln T(M,E)=
-\sum_0^n(-1)^qq\left.{d\zeta_{\Delta_q^{E}}(s)\over ds}\right\vert_{s=0}\no$$
\par
Next we turn to lens spaces.
For
general background on lens spaces cf. \ref{13,14} and references
therein---briefly, a lens space  can be constructed as follows: Take an
odd dimensional sphere $S^{2n-1}$, considered as a subset of ${\bf C}^n$,
on which a finite cyclic group of rotations $G$, say, acts. The quotient
$S^{2n-1}/G$ of the sphere under this action is a lens space. More
precisely, suppose that $G$ is of order $p$, $(z_1,\ldots,z_n)\in {\bf C}^n$
and the group action takes the form
$$\eqalign{{}&
(z_1,\ldots,z_n)\longmapsto (\exp(2\pi i q_1/p)z_1,\ldots,\exp(2\pi i
q_n/p)z_n)\cr
{}&
\hbox{with }q_1,\dots,q_n\quad\hbox{integers relatively prime to }p\cr} \no$$
then  the quotient $S^{2n-1}/G$ is a lens space often denoted by
$L(p;q_1,\dots,q_n)$. A formula
for the torsion of these spaces was first worked out by Ray \ref{2}.
We wish to focus on the situation that obtains when $n=2$ and $G$ is the
group ${\bf Z}_p\equiv {\bf Z}/p{\bf Z}$. For the most part we shall
deal with the lens space $L(2;1,1)$ which, for simplicity, we shall denote
by $L(p)$; we shall also use the notation $L(p;q)$ to denote the Lens space
$L(p;1,q)$. In passing we note that
when $p=2$ we have $L(2)={\bf R}P^3\simeq SO(3)$.
\par
The group action above defines a representation $V$, say, of
$\pi_1(L(p))$
and also determines a flat  bundle $F=(V\times S^3)/Z_p$, over $L(p)$. It is
the
torsion of this $F$ over $L(p)$ with which we are concerned here. Using
zeta functions the torsion of these lens spaces is therefore given by
$$\ln T(L(p),F)=
-\sum_0^3(-1)^qq\left.{d\zeta_{\Delta_q^{F}}(s)\over ds}\right\vert_{s=0}\no$$
\par
As an aid to the calculation of $\ln T(L(p),F)$ it is useful to introduce the
notation
$$\tau(p,s)=-\sum_0^3(-1)^qq\zeta_{\Delta_q^{F}}(s)\qquad
                 T(p)=T(L(p),F)\no$$
For $\tau(p,s)$ itself we now have
$$\tau(p,s)=\zeta_{\Delta_1^{F}}(s)-2\zeta_{\Delta_2^{F}}(s)+
3\zeta_{\Delta_3^{F}}(s)=3\zeta_{\Delta_0^{F}}(s)-\zeta_{\Delta_1^{F}}(s),\;
\hbox{using Poincar\'e duality}\no$$
Making use of the triviality of the kernels of $\Delta_q^{F}$  we further
obtain \ref{2} the formula
$$\tau(p,s)=2\zeta_{{\dstar d}_0}(s)-\zeta_{{\dstar d}_1}(s)\no$$
For the individual zeta functions we denote the eigenvalues and their
degeneracies by $\lambda_n(q,p)$ and $\Gamma_n(q,p)$ respectively giving
the expressions
$$\zeta_{{\dstar d}_0}(s)=\sum_n{\Gamma_n(0,p)\over
\lambda_n^s(0,p)},\qquad
\zeta_{{\dstar d}_1}(s)=\sum_n{\Gamma_n(1,p)\over
\lambda_n^s(1,p)}\no$$
It remains to compute these eigenvalues and degeneracies cf. \ref{15}.
The eigenvalues are
$$\lambda_n(0,p)=n(n+2),\qquad
           \lambda_n(1,p)=(n+1)^2,\;n=1,2,\dots\no$$
\par
To calculate the degeneracies is more difficult;  we make use of the
fact that $S^3$ is a group manifold and proceed as follows: Consider the
Laplacians $\dstar d_q$   on $S^3$, and $\dstar d_q^{F}$ on  $L(p)$
also, if $\lambda$ is an eigenvalue, denote the corresponding
eigenspaces by $\Lambda_q(\lambda)$ and $\Lambda^{F}_q(\lambda)$
respectively. Let
$$v(z)\in \Lambda_q(\lambda),\;\hbox{with }z\in S^3\subset {\bf C}^2,
\; \hbox{and }g\in {\bf  Z}_p,\;
\hbox{where }g\equiv\exp[2\pi i j/p],\;0\le j\le (p-1)\no$$
 The element $g$ acts on $v(z)$ to give $g\cdot v(z)$ where
$$g\cdot v(z)=v(gz)\quad\hbox{where }
gz=(\exp[2\pi i j/p]z_1,\exp[2\pi i j/p]z_2)\no$$
The above definitions allow us to define the projection $P(\lambda)$ on
$\Lambda_q(\lambda)$ by
$$P(\lambda)v={1\over p}\sum_{g\in {\bf Z}_p}\exp[-2\pi i j/p]g\cdot v\no$$
Evidently $[P(\lambda),\dstar d_q]=0$
and  so $P(\lambda)$ projects the space $\Lambda_q(\lambda)$ onto the space
$\Lambda_q^{F}(\lambda)$. Finally this means that we obtain a formula
for the degeneracy $\Gamma_n(q,p)$, namely
$$\Gamma_n(q,p)=
\tr\left(\left.P\right\vert_{\Lambda_q^{F}(\lambda)}\right)
={1\over p}\sum_{j=0}^{(p-1)}\exp[-2\pi i
j/p]\,\tr\left(\left.g\right\vert_{\Lambda_q^{F}(\lambda)}\right)\no$$
To actually apply this formula we now add in the fact that $S^3$ is the
group manifold for $SU(2)$. The Peter--Weyl theorem tells us, in this
case where all representations are self-conjugate, that
$$L^2(S^3)=L^2(SU(2))={\textstyle
\bigoplus\limits_{\mu}}\,\,c_{\mu}D_{\mu}=
{\textstyle \bigoplus\limits_{\mu}}\,\,D_\mu\otimes D_\mu\no$$
where $c_{\mu}$ measures the multiplicity of the representation $\mu$
which must therefore be $\dim D_\mu$.
But Hodge theory gives us the alternative decomposition
$$L^2(S^3)={\textstyle
\bigoplus\limits_{\lambda}}\,\,\Lambda_0(\lambda)\no$$
In addition the Casimir operator for $SU(2)$ is a multiple of the Laplacian
and, if the
 representation label $\mu$ is taken to be the usual half-integer $j$,  then
we know that this Casimir has eigenvalues $j(j+1)$, and also that
$\dim D_j=2j+1$.
These facts identify the Laplacian $\Delta_0=\dstar d_0$ as four times the
Casimir and identify $\Lambda_0(\lambda)$ as $\dim D_j$
copes of $D_j$. Thus if we set $n=2j$, so that $n$ is always integral,
then we have the degeneracy formula
$$\Gamma_n(0,p)={(n+1)\over p}\sum_{j=0}^{(p-1)}\exp[-2\pi i
j/p]\,\chi^{n/2}(2\pi j/p)\no$$
where $\chi^j(\theta)$ denotes the $SU(2)$ character, on $D_j$, for rotation
through the angle $\theta$; i.e.
$$\chi^j(\theta)={\sin((2j+1)\theta)\over\sin(\theta)}\no$$
Hence our explicit degeneracy formula for $0$-forms on $L(p)$ is
$$\Gamma_n(0,p)={(n+1)\over p}\sum_{j=0}^{(p-1)}\exp[-2\pi i
j/p]\,{\sin(2\pi(n+1)j/p)\over\sin(2\pi j/p)}\no$$
\par
We now have to find the analogous formula for the $1$-forms.
The formula that results is
$$\Gamma_n(1,p)={1\over p}\sum_{j=0}^{(p-1)}\exp[-2\pi i
j/p]\left\{n\chi^{(n+1)/2}(2\pi j/p)+(n+2)\chi^{(n-1)/2}(2\pi j/p)\right\}\no$$
\par
To simplify the notation we introduce the \lq $p$-averaged character'
$\charav{j}$
which we define by
$$\charav{j}={1\over p}\sum_{j=0}^{(p-1)}\exp[-2\pi i
j/p]\,\chi^j(2 pi j/p)\no$$
Finally this gives us a concrete expression for  $\tau(p,s)$, i.e.
$$\eqalign{\tau(p,s)&=
\sum_n\left\{{n\charav{(n+1)/2}+(n+2)\charav{(n-1)/2}\over
(n+1)^{2s}}-{2(n+1)\charav{n/2}\over \{n(n+2)\}^{2s}}\right\}\cr
 {}&=\tau_+(p,s)-\tau_-(p,s)\cr}\no$$
 with the obvious definition for $\tau_+(p,s)$ and $\tau_-(p,s)$.
\par
To actually compute the  torsion we need to
be able to evaluate these  $p$-averaged characters. This is a somewhat
non-trivial combinatorial task and it is
necessary to divide $n$ up into its conjugacy classes mod $p$ by
writing
$n=pk-j,\;k\in{\bf Z},\, j=0,1,\ldots,(p-1)$
We eventually discover that
\eqlabel{\charformula}
$$\charav{(pk-j)/2}=
\cases{
       \cases{k&for $j=0,2,\ldots,(p-1)$\cr
              k&for $j=1$\cr
              (k-1)&for $j=3,5,\ldots,(p-2)$\cr}
             &if $p$ is odd\cr
           \strut&\null\cr
      \cases{0&for $j=0,2,\dots,(p-2)$ \cr
             2k&for $j=1$\cr
             (2k-1)&for $j=3,5,\ldots,(p-1)$\cr}
             &if $p$ is even\cr}  \no$$

We must now construct the analytic continuation of the
series for $\tau(p,s)$. For the  details we refer the reader to
\ref{15}. We shall just describe here the $p=2$ case.
\beginsection{The Analytic Continuation}
The series to be continued is
$$\tau(p,s)=\sum_n\left\{{n\charav{(n+1)/2}+(n+2)\charav{(n-1)/2}\over
(n+1)^{2s}}-{2(n+1)\charav{n/2}\over \{n(n+2)\}^{2s}}\right\}
 \no$$
and it already converges for $\hbox{Re}\, s>3/2$; however a calculation of
the torsion requires us to work at $s=0$, hence we see the need for, and the
extent of, the continuation.
\par
With $p=2$ we have
$$\tau(2,s)=\sum_n\left\{{2(n+1)\left\langle\chi^{n/2}
\right\rangle_2\over \left\{n(n+2)\right\}^s}-
{n\left\langle\chi^{(n+1)/2}\right\rangle_2+
(n+2)\left\langle\chi^{(n-1)/2}\right\rangle_2
\over(n+1)^{2s}}\right\}\no$$
But using \docref{charformula} we find that
$$\eqalign{\left\langle\chi^{(n+1)/2}\right\rangle_2&=
           \left\langle\chi^{(2k-j+1)/2}\right\rangle_2,\quad (n=2k-j)\cr
        \strut&{}\cr
        {}&=\cases{0,& $j=1$\cr
                    2k+2,& $j=0$\cr}
             \quad\equiv\cases{0,& $n$ odd\cr
                    2k+2,& $n$ even\cr}\cr}\no$$

Similarly
$$\eqalign{\left\langle\chi^{(n-1)/2}\right\rangle_2&=
\left\langle\chi^{(2k-j)/2}\right\rangle_2,\quad (n=2k-j)\cr
        \strut&{}\cr
        {}&=\cases{2k,& $j=1$\cr
                    0,& $j=0$\cr}
             \quad\equiv\cases{(n+1),& $n$ odd\cr
                    0,& $n$ even\cr}\cr}\no$$
Thus $\tau(2,s)$ becomes
$$\tau(2,s)=\tau_+(2,s)-\tau_-(2,s)
                    =\sum_{n\;{\rm odd}}{
2(n+1)^2\over\left\{n(n+2)\right\}^s}
-\sum_{n\;{\rm even}}{2n(n+2)\over(n+1)^{2s}}\no$$
Setting $n=(2m-1)$ in $\tau_+(2,s)$ and $n=2m$ in $\tau_-(2,s)$
we have
$$\eqalign{\tau(2,s)&=\sum_{m=1}^\infty{8 m^2\over(4m^2-
1)^s}-\sum_{m=0}^\infty{2m(2m+2)\over (2m+1)^{2s}}\cr
{}&=\sum_{m=1}^\infty{8 m^2\over(4m^2-
1)^s}-\sum_{m=0}^\infty{1\over (2m+1)^{(2s-2)}}
+\sum_{m=0}^\infty{1\over (2m+1)^{2s}}\cr}\no$$
Now if $\zeta(s)$ is the usual Riemann zeta function we can use the fact that
$$\sum_{n=1,3,5,\dots}{1\over n^s}=(1-2^{-s})\zeta(s)\no$$
then we get
\eqlabel{\zetaexpress}
$$\tau(2,s)=\sum_{m=1}^\infty{8 m^2\over(4m^2-
1)^s}-2(1-2^{-(2s-2)})\zeta(2s-2)+2(1-2^{-2s})\zeta(2s)\no$$
The only term in \docref{zetaexpress} without a well defined continuation
 is the first term.
To this end we define the quantity $A(m,s)$ by
$$\eqalign{A(m,s)&={4m^2\over(4m^2-1)^s}={4m^2\over (4m^2)^s}
                      \left(1-{1\over4m^2}\right)^{-s}
                ={1\over(4m^2)^{(s-
                                    1)}}\left\{1+{s\over4m^2}+\cdots\right\}\cr
                {}&={1\over(4m^2)^{(s-1)}}+{s\over(4m^2)^s}+R(m,s),
                                        \qquad(\hbox{def. of }R(m,s))\cr}\no$$
So that the remainder term $R(m,s)$ is given by
$$\eqalign{R(m,s)&=A(m,s)-{1\over(4m^2)^{(s-1)}} -{s\over(4m^2)^s}\cr
                 &={4m^2\over(4m^2-1)^s}-{1\over(4m^2)^{(s-1)}}-
                                                  {s\over(4m^2)^s}\cr}\no$$
\par
The definition of the remainder term is chosen to ensure that
$$\left\vert R(m,s)\right\vert\le{(\ln m)^\alpha\over m^2}\no$$
and this has the vital consequence that the operations $d/ds$ (at $s=0$) and
$\sum_m$ {\it commute } when applied to $R(m,s)$.
\par
Returning to $\tau(s,2)$ itself we have
$$\eqalign{\tau(2,s)&=2\sum_{m=1}^\infty A(m,s)
-2(1-2^{-(2s-2)})\zeta(2s-2)+2(1-2^{-2s})\zeta(2s)\cr
\Rightarrow \tau(2,s)&=2\sum_{m=1}^\infty{1\over(4m^2)^{(s-1)}}+
2s\sum_{m=1}^\infty{1\over(4m^2)^s}+2\sum_{m=1}^\infty R(m,s)\cr
&-2(1-2^{-(2s-2)})\zeta(2s-2)+2(1-2^{-2s})\zeta(2s)\cr}\no$$
Defining
$$R(s)=\sum_{m=0}^\infty R(m,s)\no$$
gives a series  for $R(s)$ which is guaranteed to be convergent and
the analytic
continuation is now complete; thus we can now take the final step which is to
differentiate and obtain the torsion  $T(2)$.
The result that we get is that
$$\ln T(2)={d\tau(2,0)\over ds}=28\zeta^\prime(-2)
+2(1+\ln4)\zeta(0)+2R^\prime(0)\no$$
But it is easy to check that
$\zeta(0)=-1/2$ and $\zeta^\prime(-2)=-
{\zeta(3)/4\pi^2}$ and by our remark above concerning the motive for
our choice of definition
for $R(m,s)$ we have
$$\eqalign{R^\prime(0)&=
{d\over ds}\sum_m\left. R(m,s)\right\vert_{s=0}\cr
\Rightarrow R^\prime(0)&=\sum_m\left.{d R(m,s)\over ds}\right\vert_{s=0}\cr
                       &=\sum_m\left[4m^2\left\{\ln(4m^2)-
\ln(4m^2-1)\right\}-1\right]=-
\sum_m\left[4m^2\ln(1-1/4m^2)+1\right]\cr}\no$$
Hence
$$\ln T(2)=-{7\over\pi^2}\zeta(3)-1-2\ln(2)-
2\sum_m\left[4m^2\ln(1-1/4m^2)+1\right]\no$$
However the series for $R^\prime(0)$ can be expressed as a trigonometric
integral cf. \ref{15}. In fact we have
$$\sum_{m=1}^\infty\left[4m^2\ln(1-1/4m^2)+1\right]=-{1\over2}+
{4\over\pi^2}\int_0^{\pi/2}dz\,z^2\cot(z)\no$$
which means that
\eqlabel{\tortwoexpr}
$$\ln T(2)=-{7\over\pi^2}\zeta(3)-2\ln(2)-
{8\over\pi^2}\int_0^{\pi/2}dz\,z^2\cot(z)\no$$
This formula \docref{tortwoexpr} above for $T(2)$ can be pushed even further;
by using  Ray's expression \ref{2} for the torsion  we can
deduce that
\eqlabel{\rayexpr}
$$\ln T(p)=-{4\over p}\sum_{j=1}^{(p-1)}\sum_{k=1}^p\cos({2jk\pi\over
p})\ln(2\sin({2k\pi\over p}))\exp[{2k\pi i\over
p}]=-4\ln(2\sin({\pi\over p}))\no$$
which, for $p=2$, becomes simply
$$\ln T(2)=-4\ln(2)\no$$
Hence we straightaway have that
$$-4\ln(2)=-{7\over\pi^2}\zeta(3)-2\ln(2)-
{8\over\pi^2}\int_0^{\pi/2}dz\,z^2\cot(z)\no$$
Or
$$\zeta(3)={2\pi^2\over7}\ln(2)-{8\over7}\int_0^{\pi/2}dz\,z^2\cot(z)\no$$
in other words our computation of the torsion has given us a formula for
$\zeta(3)$.
\par
In \ref{15} we construct the continuation for arbitrary $p$ but here we
limit ourselves to quoting the torsion formula for $p$ odd which (recall that
$\ln T(p)=-4\ln(2\sin({\pi/ p}))$) is
$$\eqalign{\ln T(p)=&-
{(p^3-1)\over p}{\zeta(3)\over\pi^2}-{{2\over {\pi }}
\,\left( p-2 \right) \, \int _{0}^{{{\pi }\over p}}
         dz\,z\,\cot (z)  } +
  {{2\over {\pi }}\,\left( p-2  \right) \, \int _{0}
          ^{{{\left( p-1  \right) \,\pi }\over p}}dz\,z\,\cot (z) }\cr
 &- {{2\,p\over {{{\pi }^2}}}\, \int _{0}^{{{\pi }\over p}}
         dz\,{z^2}\,\cot (z)  }
-{{2\,p\over {{{\pi }^2}}}\, \int _{0}^{{{\left( p-1 \right) \,\pi }\over p}}
         dz\,{z^2}\,\cot (z)  } -{{4\,\ln (2\,\sin ({{\pi }\over p}))}\over p}
\cr
 &+{16\over {\pi }}\, \sum_{l = 1}^{(p-3)/2}
         l\, \int _{0}^{{{2\,l\,\pi }\over p}}dz\,z\,\cot (z)
 -{{4\,p\over
    {{{\pi }^2}}}\, \sum_{l = 1}^{(p-3)/2}
         \int _{0}^{{{2\,l\,\pi }\over p}}dz\,{z^2}\,\cot (z)  }\cr
&-{{4\over p}\, \sum_{l = 1}^{{(p-3)/ 2}}
         4\,{l^2}\,\ln (2\,\sin ({{2\,l\,\pi }\over p}))  },\qquad
\hbox{for $p$ odd}\cr}\no$$
\beginsection{Concluding remarks}
These formulae have yet to be elucidated further.
\par
A thought provoking fact is
that $\zeta(3)$ occurs in a recent paper of Witten \ref{16} where, after
multiplication by a known constant, it gives the volume of the
symplectic space of flat connections over a {\it non-orientable} Riemann
surface. The corresponding calculation for  {\it orientable} surfaces
(where the volume element is a rational cohomology class)
allows a cohomological rederivation of the irrationality of $\zeta(2),\,
\zeta(4),\dots$. This paper also involves the torsion but in two dimensions
rather than three.
The proof that $\zeta(3)$ is irrational was only obtained in 1978 cf. \ref{17}
and the rationality of $\zeta(5),\,\zeta(7),\ldots$ is at present open.
\par
Our technique, applied in five dimensions instead of three would yield
formulae for $\zeta(5)$ but their nature is as yet unclear.
\par
Further interesting
results are that the  $T(p)$ is trivial (i.e unity) when $p=6$; and that if we
work with $L(p,q)$
rather than $L(p)$ then the only other  three dimensional lens spaces for
which the torsion is trivial are $L(10,3)$ and $L(12,5)$. The large $p$
behaviour of the determinants is also computable: $ T(p)$ grows as
$p^4/(2\pi)^4+p^2/6(2\pi)^2$ for large $p$, while the determinants
grow much faster than quartically.
\par\vskip\baselineskip
\centerline{\bf References}
\vskip0.5\baselineskip
{ \par \noindent \par \hangindent \parindent \indent
\hbox to 0pt {\hss \fam \bffam \tenbf 1.\kern .5em }
\ignorespaces
  Franz W., {\accent "7F U}ber die Torsion einer {\accent "7F U}berdeckung,
J. Reine Angew. Math.,
{\fam \bffam \tenbf 173}, 245--254, (1935).
\par \vskip -0.8\baselineskip \noindent }
{ \par \noindent \par \hangindent \parindent \indent
\hbox to 0pt {\hss \fam \bffam \tenbf 2.\kern .5em }
\ignorespaces
  Ray D. B., Reidemeister torsion and the Laplacian on lens spaces,
Adv. in Math.,
{\fam \bffam \tenbf 4}, 109--126, (1970).
\par \vskip -0.8\baselineskip \noindent }
{ \par \noindent \par \hangindent \parindent \indent
\hbox to 0pt {\hss \fam \bffam \tenbf 3.\kern .5em }
\ignorespaces
  Ray D. B. and Singer I. M.,
R-torsion and the Laplacian on Riemannian manifolds, Adv. in Math.,
{\fam \bffam \tenbf 7}, 145--201, (1971).
\par \vskip -0.8\baselineskip \noindent }
{ \par \noindent \par \hangindent \parindent \indent
\hbox to 0pt {\hss \fam \bffam \tenbf 4.\kern .5em }
\ignorespaces
  Ray D. B. and Singer I. M.,
Analytic Torsion for complex manifolds, Ann. Math.,
{\fam \bffam \tenbf 98}, 154--177, (1973).
\par \vskip -0.8\baselineskip \noindent }
{ \par \noindent \par \hangindent \parindent \indent
\hbox to 0pt {\hss \fam \bffam \tenbf 5.\kern .5em }
\ignorespaces
  Cheeger J., Analytic torsion and the heat equation, Ann. Math.,
{\fam \bffam \tenbf 109}, 259--322, (1979).
\par \vskip -0.8\baselineskip \noindent }
{ \par \noindent \par \hangindent \parindent \indent
\hbox to 0pt {\hss \fam \bffam \tenbf 6.\kern .5em }
\ignorespaces
  M{\accent "7F u}ller W.,
Analytic torsion and the R-torsion of Riemannian manifolds, Adv. Math.,
{\fam \bffam \tenbf 28}, 233--, (1978).
\par \vskip -0.8\baselineskip \noindent }
{ \par \noindent \par \hangindent \parindent \indent
\hbox to 0pt {\hss \fam \bffam \tenbf 7.\kern .5em }
\ignorespaces
  Witten E.,
{\fam \itfam \tenit Quantum field theory and the Jones polynomial},
I. A. M. P. Congress, Swansea, 1988, {edited by: Davies I., Simon B. and Truman
A.}, Institute of Physics, (1989).
 \par \vskip -0.8\baselineskip \noindent }
{ \par \noindent \par \hangindent \parindent \indent
\hbox to 0pt {\hss \fam \bffam \tenbf 8.\kern .5em }
\ignorespaces
  Witten E.,
Quantum field theory and the Jones polynomial, Commun. Math. Phys.,
{\fam \bffam \tenbf 121}, 351--400, (1989).
\par \vskip -0.8\baselineskip \noindent }
{ \par \noindent \par \hangindent \parindent \indent
\hbox to 0pt {\hss \fam \bffam \tenbf 9.\kern .5em }
\ignorespaces
  Nash C.,
{\fam \itfam \tenit Differential Topology and Quantum Field Theory},
Academic Press, (1991).
 \par \vskip -0.8\baselineskip \noindent }
{ \par \noindent \par \hangindent \parindent \indent
\hbox to 0pt {\hss \fam \bffam \tenbf 10.\kern .5em }
\ignorespaces
  Birmingham D., Blau M., Rakowski M. and Thompson G.,
Topological field theory, Phys. Rep.,
{\fam \bffam \tenbf 209}, 129--340, (1991).
\par \vskip -0.8\baselineskip \noindent }
{ \par \noindent \par \hangindent \parindent \indent
\hbox to 0pt {\hss \fam \bffam \tenbf 11.\kern .5em }
\ignorespaces
  Batalin I. A. and Vilkovisky G. A.,
Quantisation of Gauge Theories with linearly independent generators,
Phys. Rev.,
{\fam \bffam \tenbf D28}, 2567--, (1983).
\par \vskip -0.8\baselineskip \noindent }
{ \par \noindent \par \hangindent \parindent \indent
\hbox to 0pt {\hss \fam \bffam \tenbf 12.\kern .5em }
\ignorespaces
  Batalin I. A. and Vilkovisky G. A.,
Existence theorem for gauge algebras, Jour. Math. Phys.,
{\fam \bffam \tenbf 26}, 172--, (1985).
\par \vskip -0.8\baselineskip \noindent }
{ \par \noindent \par \hangindent \parindent \indent
\hbox to 0pt {\hss \fam \bffam \tenbf 13.\kern .5em }
\ignorespaces
  Rolfsen D.,
{\fam \itfam \tenit Knots and Links}, Publish or Perish, (1976).
 \par \vskip -0.8\baselineskip \noindent }
{ \par \noindent \par \hangindent \parindent \indent
\hbox to 0pt {\hss \fam \bffam \tenbf 14.\kern .5em }
\ignorespaces
  Bott R. and Tu L. W.,
{\fam \itfam \tenit Differential Forms in Algebraic Topology},
Springer-Verlag, New York, (1982).
 \par \vskip -0.8\baselineskip \noindent }
{ \par \noindent \par \hangindent \parindent \indent
\hbox to 0pt {\hss \fam \bffam \tenbf 15.\kern .5em }
\ignorespaces
  Nash C. and O' Connor D. J., in preparation.
\par \vskip -0.8\baselineskip \noindent }
{ \par \noindent \par \hangindent \parindent \indent
\hbox to 0pt {\hss \fam \bffam \tenbf 16.\kern .5em }
\ignorespaces
  Witten E.,
On quantum gauge theories in two dimensions, Commun. Math. Phys.,
{\fam \bffam \tenbf }, 153--209, (1991).
\par \vskip -0.8\baselineskip \noindent }
{ \par \noindent \par \hangindent \parindent \indent
\hbox to 0pt {\hss \fam \bffam \tenbf 17.\kern .5em }
\ignorespaces
  Poorten Alfred van der,
A proof that Euler missed.... Ap{\accent 19 e}ry's proof of the
irrationality of $\zeta (3)$. An informal report.,
The Mathematical Intelligencer,
{\fam \bffam \tenbf 1}, 195--203, (1979).
\par \vskip -0.8\baselineskip \noindent
}\bye